\documentclass[twocolumn,english,aps,superscriptaddress]{revtex4}

\usepackage{amssymb}
\usepackage{amsmath}
\usepackage{colordvi}
\usepackage[colorlinks]{hyperref}
\usepackage{amsthm}
\usepackage{subeqnarray}
\usepackage{cases}
\usepackage{enumerate}
\usepackage{graphicx}
\usepackage{epsfig}
\usepackage{epstopdf}
\usepackage{subfigure}
\usepackage{color}
\usepackage{url}
\usepackage{verbatim}
\usepackage{mathrsfs}

\usepackage{braket}
\newcommand{\vecbb}[1]{\mathbf{#1}}

\begin{document}

\title{Quantum light from lossy semiconductor Rydberg excitons}
\author{Valentin Walther}
\email{valentin.walther@cfa.harvard.edu}
\affiliation{ITAMP, Harvard-Smithsonian Center for Astrophysics, Cambridge, Massachusetts 02138, USA}
\affiliation{Department of Physics, Harvard University, Cambridge, Massachusetts 02138, USA}
\author{Anders S. S\o{}rensen}
\affiliation{Center  for  Hybrid  Quantum  Networks  (Hy-Q),  The  Niels  Bohr Institute,  University  of  Copenhagen,  DK-2100  Copenhagen  \O{}Ø,  Denmark}

\begin{abstract}
 The emergence of photonic quantum correlations is typically associated with emitters strongly coupled to a photonic mode. Here, we show that semiconductor Rydberg excitons, which are only weakly coupled to a free-space light mode can produce strongly antibunched fields, i.e. quantum light. This effect is fueled by micron-scale excitation blockade between Rydberg excitons inducing pair-wise polariton scattering events. Photons incident on an exciton resonance are scatted into blue- and red-detuned pairs, which enjoy relative protection from absorption and thus dominate the transmitted light. We demonstrate that this effect persists in the presence of additional phonon coupling, strong non-radiative decay and across a wide range of experimental parameters. Our results pave the way for the observation of quantum statistics from weakly coupled semiconductor excitons.
\end{abstract}

\maketitle

\begin{figure}[t]
\begin{center}
 \includegraphics[height=.35\textwidth]{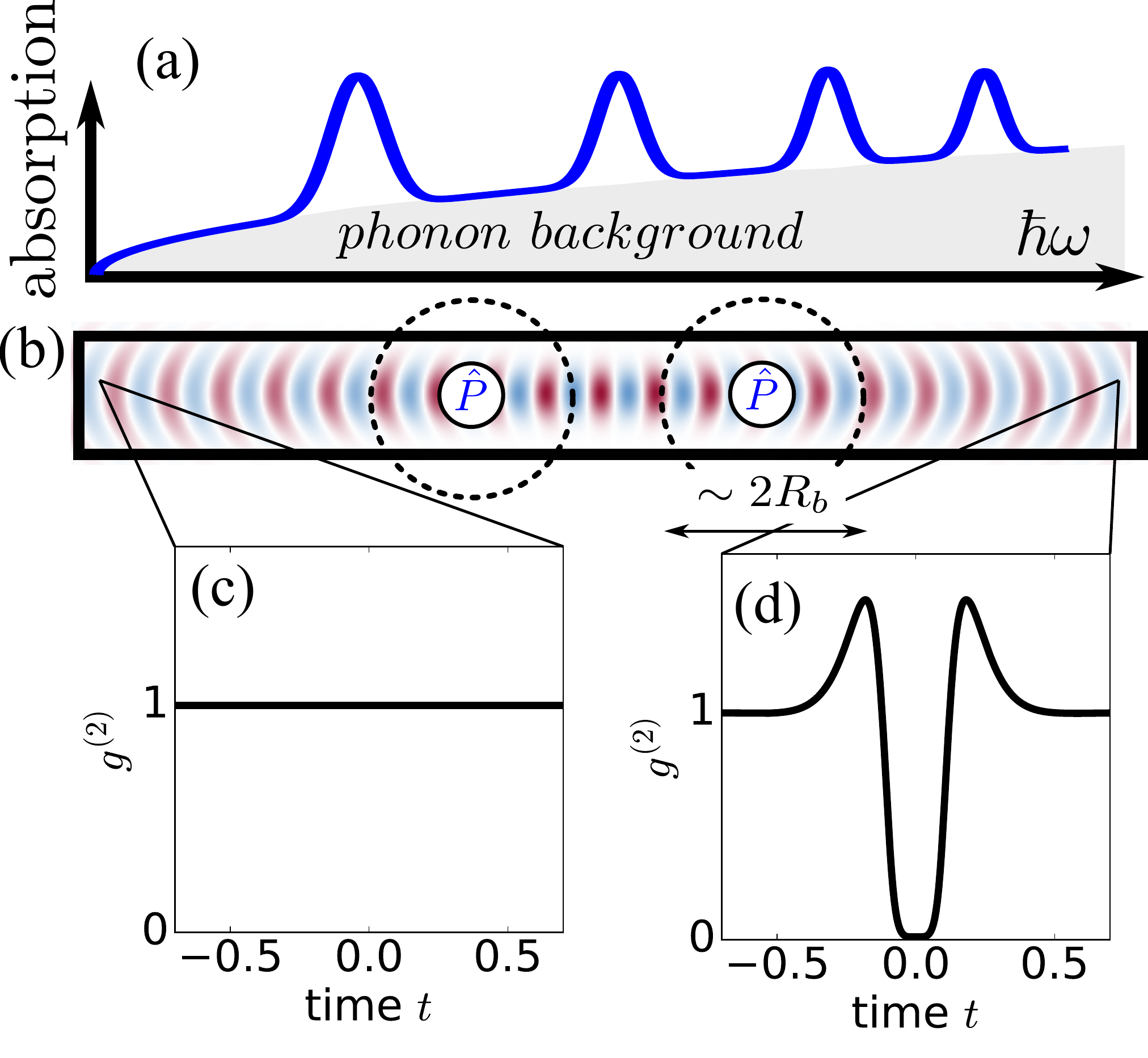}
\end{center}
\caption{The spectrum of Rydberg excitons, as in cuprous oxide, is often superimposed on a broad phonon-assisted absorptive background (a). Rydberg excitons ($\hat{P}$), as created by near-resonant light, blockade each other over a distance of $R_b$ (b). This correlated dynamics converts a weak coherent beam at the input (c) into a highly nonclassical photonic state (d), as evidenced by the second-order correlation function $g^{(2)}(0) \approx 0$ (for $L \approx 2.4$, $\beta = 10$, $R_b = 0.1$).}
\label{fig1}
\end{figure}

Semiconductor excitons, bound electron-hole pairs, have been at the forefront of semiconductor research ever since their discovery in the 1950s. One of their central features is their optical activity \cite{gross1956} and hence their innate ability to interface with light. These capabilities have been demonstrated in a number of observations including polariton formation \cite{weisbuch1992}, parametric scattering \cite{savvidis2000_angle, savvidis2000_asymmetric}, four-wave mixing \cite{ciuti2000, ciuti2001}, optical bistability \cite{baas2004, amo2010} and polariton lasing \cite{kim2016, bajoni2012}. Despite intense recent efforts \cite{munjoz2019, delteil2019}, pushing these effects down to the level of individual photons has proven difficult. The observation of highly excited Rydberg states of excitons \cite{kazimierczuk2014giant} promises access to strong interactions \cite{walther2018interactions}, an ingredient that may prove valuable for producing quantum light using the effect of Rydberg blockade, as pioneered in atomic gases with great success \cite{peyronel2012, saffman2002, dudin2012}. 

In Rydberg blockade, the presence of an excitation inside the medium inhibits the excitation of another Rydberg state within the so-called blockade radius $R_b$ which in semiconductors can be several microns large \cite{walther2020}. The highest principal quantum number of $n \sim 30$ and largest associated blockade radii were observed in the so-called yellow series of Cu$_2$O \cite{kazimierczuk2014giant, steinhauer2020, versteegh2021giant}. These exciton states feature only relatively weak dipole transition elements, as the valence and conduction band have the same parity \cite{elliott1957}. In addition, the exciton series is superimposed by a phonon-assisted absorptive background, which can account for more than half of the absorption on a Rydberg exciton resonance (Fig.~\ref{fig1}(a)). While both factors, in general, counteract the buildup of quantum correlations, we here describe conditions under which quantum light can still be observed from lossy semiconductor Rydberg excitons. 

In particular, we describe the transmission of light through a crystal with Rydberg excitons, e.g. Cu$_2$O, and show that the transmitted light can show quantum properties. This result is reminiscent of similar effects observed in atoms coupled to optical waveguides \cite{prasad2020, mahmoodian2018}. We further establish a formal link between Rydberg exciton transmission and discrete scatterers coupled to waveguides. We expect that this connection will be a fruitful link for the exploration of excitonic Rydberg states, where quantum states of light have so far not been observed.

We consider a cuprous oxide crystal of thickness $L$ which is illuminated by a light beam detuned by a frequency $\Delta = \omega - \omega_x$ from one of the discrete exciton resonances $\omega_x$ (Fig.~\ref{fig1}). Focusing on a paraxial light beam under orthogonal illumination, we consider photons in a single transverse mode, which can be described by the bosonic operator $\hat{\mathcal{E}}(\vecbb{r})$ that denotes the slowly-varying electric-field envelope of the electromagnetic field and yields the photon density operator $\hat{\mathcal{E}}^\dagger\hat{\mathcal{E}}$. In the energy range of the Rydberg series, the field propagates at the group velocity $v_g$ \cite{scully1997}. During its propagation, it near resonantly couples to excitons in the semiconductor, described by the bosonic operator $\hat{\mathcal{P}}(\vecbb{r})$. However, it also couples to a broad absorptive phonon-assisted resonance that underlies the Rydberg exciton spectrum in Cu$_2$O \cite{schoene2017}. This background can be integrated out \cite{waltherphonon2020} to effectively give a finite absorption rate $\gamma_\text{bg}$ in the semiconductor. The light propagation is hence described by the effective Hamiltonian 
\begin{equation}
\begin{aligned}
 \hat{\mathcal{H}}_0/\hbar = -&\Delta \int d\vecbb{r} \hat{\mathcal{P}}^\dagger(\vecbb{r}) \hat{\mathcal{P}}(\vecbb{r}) 
 + g \int d\vecbb{r} \hat{P}^\dagger(\vecbb{r}) \hat{\mathcal{E}}(\vecbb{r})  + \text{h.c.} \\
 +& \int d \vecbb{r}\hat{\mathcal{E}}^\dagger(\vecbb{r}) \left( -i v_g \partial_z + \frac{\hbar}{2 m_\perp} \nabla^2_\perp \right)  \hat{\mathcal{E}}(\vecbb{r}) \\
 -&i \gamma_\text{bg} \int d\vecbb{r}\hat{\mathcal{E}}^\dagger(\vecbb{r}) \hat{\mathcal{E}}(\vecbb{r}),
\end{aligned} \label{eq:hamiltonian_unidirectional}
\end{equation}
with the light-matter coupling strength $g$ and the transverse effective mass $m_\perp = n\hbar \omega_x/(c v_g)$ with the dielectric constant $n$ and the speed of light $c$. The source of photon correlations lies in extraordinarily strong interactions between highly-excited Rydberg excitons \cite{saffman2010} due to long-ranged dipole-dipole interactions. These interactions give rise to an excitation blockade wherever they exceed the exciton linewidth $\gamma$. While the optically active excitons in Cu$_2$O have $p$-state character and are thus multiply degenerate \cite{PhysRev.124.340, schoene2016}, we employ here a simplified model with a single effective potential energy surface \cite{stanojevic2016} $\hat{\mathcal{V}} = \frac{1}{2} \int d\vecbb{r} \int d\vecbb{r}' V(|\vecbb{r}-\vecbb{r}'|) \hat{\mathcal{P}}^\dagger(\vecbb{r}) \hat{\mathcal{P}}^\dagger(\vecbb{r}') \hat{\mathcal{P}}(\vecbb{r}') \hat{\mathcal{P}}(\vecbb{r})$ with $V(R) = C_6/R^6$. The total Hamiltonian is thus $\hat{\mathcal{H}} = \hat{\mathcal{H}}_0 + \hat{\mathcal{V}}$.

We describe the quantum dynamics in the semiconductor by expanding the overall wave function into sectors of a single, two, three, and so on excitations. If the crystal is excited by a weak coherent light field, each sector is suppressed by a small amplitude $\alpha$ compared to sectors of lower excitation number. Hence, we approximate the full state by the sector for 0, 1 and 2 excitations only. In the single-excitation sector, the excitation can either reside in the photon or in the exciton, as described by the state $|\Psi(t) \rangle^{(1)} = \int d\vecbb{r} E(\vecbb{r},t) \hat{\mathcal{E}}^\dagger(\vecbb{r}) |\emptyset\rangle 
 + \int d\vecbb{r} P(\vecbb{r},t) \hat{\mathcal{P}}^\dagger(\vecbb{r}) |\emptyset\rangle$. In the doubly-excited sector, the possible combinations are: having two photons ($EE(\vecbb{r},\vecbb{r'},t)$), two excitons ($PP(\vecbb{r},\vecbb{r'},t)$) or one of each ($EP(\vecbb{r},\vecbb{r'},t)$), where the first coordinate describes the photon and the second the exciton location.

To elucidate the dynamics of the system, we apply the Schr\"odinger equation to each excitation sector $|\Psi(t) \rangle^{(n)}$. We note that loss from the $n$-excitation sector, in principle, contributes to the sectors with fewer excitations. However, due to the relative suppression with $\alpha$ such contributions are small and can be neglected, leaving separate dynamics in each sector. The noninteracting paraxial wave equations are solved by the complete set of Laguerre-Gauss and Hermite-Gauss modes, as demonstrated in Appendix~\ref{sec:mode_expansion}. For simplicity, we focus on and project into the lowest such mode, the $TEM_{00}$ mode, and discuss the general case in Appendix~\ref{sec:mode_expansion}. 

The equations of motion for a single excitation in the crystal are 
\begin{align}
 \partial_t E(z,t) &= - v_g \partial_z E(z,t) - \gamma_\text{bg} E(z,t) - i g P(z, t) \\
 \partial_t P(z,t) &= - ig E(z, t) - \frac{\Gamma}{2} P(z, t),
\end{align}
where we have defined the complex linewidth $\Gamma = \gamma - 2i \Delta$ with $\gamma$ being the combined excitonic decay (out of the mode) and dephasing rate. In the continuous-wave (cw) limit considered here and in the following, these equations are readily solved and we obtain
\begin{align}
 P(z) = - \frac{2ig}{\Gamma} E(z), \qquad E(z) = \alpha e^{- \left( \frac{2g^2}{v_g\Gamma} + \frac{\gamma_\text{bg}}{v_g} \right)z},
\end{align}
describing standard Beer-Lambert absorption. This solution reveals some important quantities, which we employ to re-scale our equations: the resonant excitonic field absorption length, $l_\text{abs} = \frac{v_g\gamma}{2g^2}$ ($z \rightarrow z/l_\text{abs}$), the associated absorption time, $t_\text{abs} = l_\text{abs}/v_g$ ($t \rightarrow t/t_\text{abs}$), the ratio of coherence and absorption lengths, $\beta = \gamma^2/(4g^2)$, and the scaled detuning $x = \Delta/\gamma$. To focus on the effects of interactions, we move to an interaction picture by factoring the linear solutions $f_0(z,z')$ out of each two-excitation wave function component $F(z,z')$ via $F(z,z') = f(z,z')f_0(z,z')$. As shown explicitly in Appendix~\ref{sec:derivation_details}, two excitations in the crystal then follow the coupled dynamics
\begin{equation} \label{eq:two_exc_dynamics}
\begin{aligned}
 \partial_R ee(R,r) &= \frac{2}{1-2ix} ee(R,r) -\frac{1}{1-2ix} ep_+(R,r) \\
 \partial_R ep_+ (R,r) &= -2\partial_r ep_- (R,r) + 4(1-2ix)\beta ee(R,r) \\
 &+ 2\left(-(1-2ix)\beta -i V_\text{eff}(r)\right) ep_+ (R,r) \\
 \partial_R ep_- (R,r) &= -2\partial_r ep_+ (R,r) -2(1-2ix) \beta ep_- (R,r) \\
 &+ \frac{2}{1-2ix} ep_- (R,r).
 \end{aligned}
\end{equation}
Here, we have switched to center-of-mass and relative coordinates, defined by $R = \frac{z+z'}{2}$ and $r = z-z'$ and use symmetrized and antisymmetrized wave functions. Note that these equations are only strictly valid in the cw-limit, where the wave function component describing two excitons, $PP$, can be eliminated exactly. The general time-dependent equations are given in Appendix~\ref{sec:derivation_details}. Special attention must be attributed to the boundary conditions of Eq.~(\ref{eq:two_exc_dynamics}), describing the state of the system when one excitation is the medium, while the other has not entered or has already exited. They are given by $ee(R,r) = 1$, $ep_+ (R,r)=2$ and $ep_- (R,r)=0$ when $z=0$ or $z'=0$. Without interactions ($V_\text{eff}(r)=0$), the solution is constant and given by the boundary conditions, showing that only the interaction populates the antisymmetric wave function component. Conveniently, these transformations render a particularly simple form of the second-order photonic correlation function in transmission $g^{(2)}(\tau) = |ee(L,L-v_g\tau)|^2$. 

\begin{figure}[h]
\begin{center}
 \includegraphics[height=.45\textwidth]{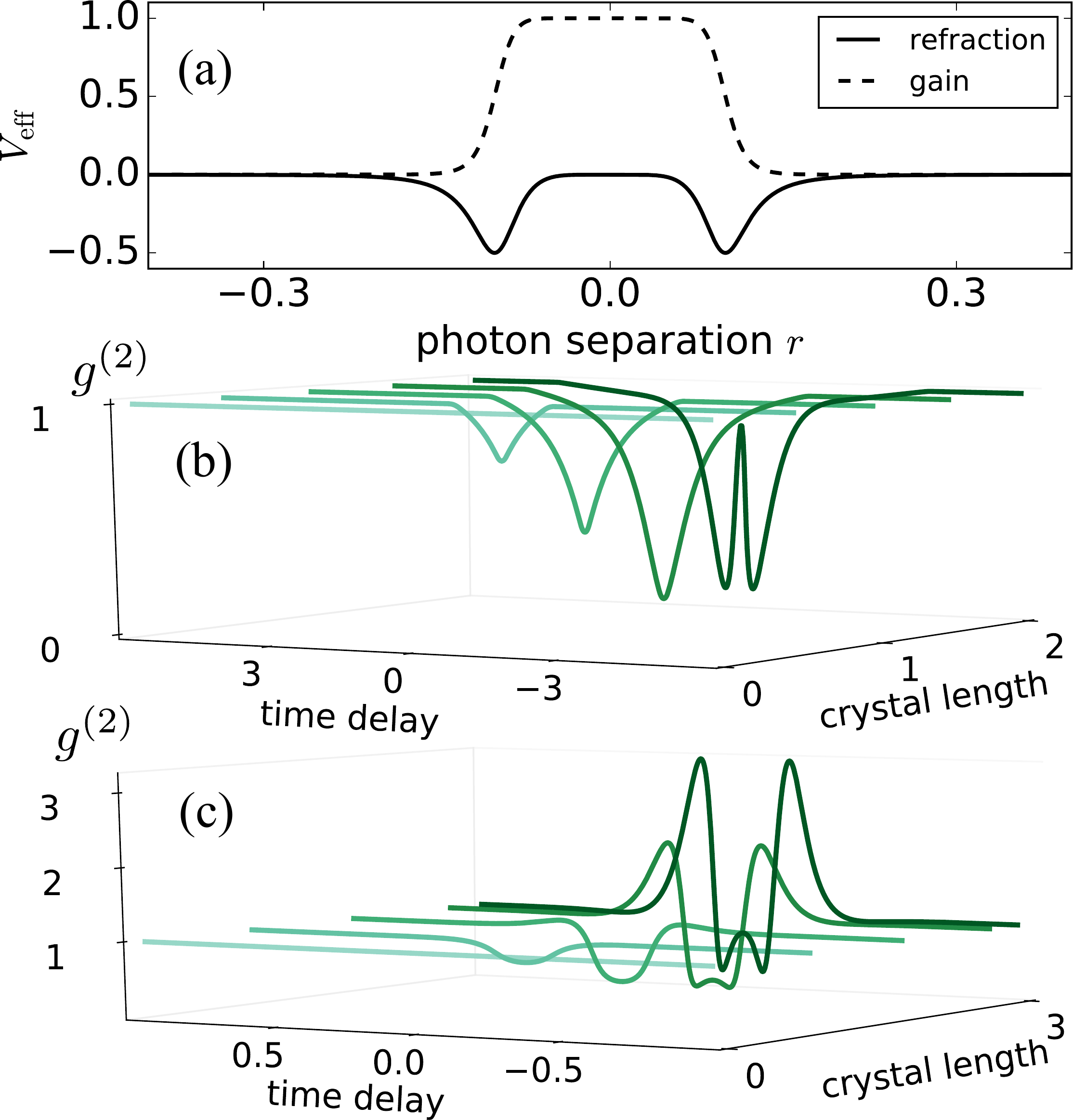}
\end{center}
\caption{During their propagation through the crystal, the excitations are subject to a Rydberg-mediated interaction potential (a), featuring nonlinear absorption/gain (imaginary part, dashed line) as well nonlinear refraction (real part, solid line). The response at short distances takes a flat-top profile, reflecting the excitation blockade, shown here for $R_b=0.1$. Low-loss conditions (b, $\beta = 1$) feature long-range photon correlations, and photon bunching at $t=0$ for long crystals, while high-loss (c, $\beta = 10$) evokes correlations on the scale of $R_b$ and \textit{finite-time} bunching for long crystals. In either case, quantum light, $g^{(2)}(0) < 0$, can be found at some crystal lengths.}
\label{fig2}
\end{figure}

Remarkably, we observe in Eq.~(\ref{eq:two_exc_dynamics}) that the problematic phonon-assisted absorption background $\gamma_\text{bg}$ can identically be scaled out of the equations, showing that the photon correlations in transmission are immune against loss from such parallel absorption pathways. Thus, we can analyze the transmission of an initially flat beam through the Cu$_2$O crystal (as illustrated in Fig.~\ref{fig1}) and interpret the emergent correlations as signatures of the semiconductor's Rydberg excitations. 

As explained above, a key role is played by the effective potential $V_\text{eff}$, which breaks the non-interacting solution and induces correlations. As derived in Appendix~\ref{sec:derivation_details}, its functional form is 
\begin{align}
 V_\text{eff}(r) = -\frac{1}{2x+i} + \frac{1}{2x - \frac{R_b^6}{r^6}+i}.
\end{align}
Fig.~\ref{fig2}(a) illustrates the effective potential for resonant excitation and repulsive exciton-exciton interactions \cite{balewski2014}. Its intrinsic length scale is the blockade radius, dividing the potential into a perturbative part at large particle separations and a non-perturbative flat part at short separations. The dominant role is played by the short-ranged plateau, representing a nonlinear reduction of absorption. Alongside this nonlinear absorption, there is a ``refractive'' effect at finite distances reflecting a change in the effective detuning caused by a Rydberg excitation. 

Strictly speaking, the photon propagation defines a boundary value problem, which needs to be solved on the domain $(z,z') \in [0,L]$. However, since the absorption length in Cu$_2$O is much longer than the blockade radius, $l_\text{abs} \gg R_\text{bl}$, the minimal build-up of correlations when the photons enter and exit the crystal can be neglected and Eq.~(\ref{eq:two_exc_dynamics}) be solved as an initial value problem.

We begin discussing the case of long coherence length ($\beta \approx 1$). This regime is characterized by relatively high coherence, where an excitation has equal coupling rate to the input mode and other channels. Fig.~\ref{fig2}(b) shows how propagation through the crystal converts an initially flat state into a nontrivial photonic correlation function: After short propagation distances, the $g^{(2)}$ function develops a dip at short time separations $\tau$. As the photons propagate further through the crystal, the dip widens and deepens, until it hits the limit of very antibunched light, $g^{(2)}(0) \approx 0$. Beyond that point, the second-order correlation function changes direction and develops a distinct maximum at $\tau=0$, while new minima emerge at finite $\tau$. In the limit of very long crystals, the correlations grow at zero time delay, producing bunched light $g^{(2)}(0) \rightarrow \infty$.

The above behavior can be understood by unraveling the scattering events in the crystal. Transforming Eq.~(\ref{eq:two_exc_dynamics}) into reciprocal space in the relative coordinate yields $i\partial_R \vec{\psi}(R) = \left( \vecbb{H_0} + \vecbb{V}  \right) \vec{\psi}(R)$, with a noninteracting diagonal matrix $\vecbb{H}_0$ and an off-diagonal interaction matrix $\vecbb{V}$. This offers a convenient perturbation expansion in $\vecbb{V}$, where the excitation pair experiences no scattering event, one scattering event, two scattering events and so on. Focusing on the lowest two orders, we obtain $\vec{\psi}(L) = \vecbb{T}(L) \vec{\psi}(0)$ with the transfer matrix $\vecbb{T}(L) = \exp \left( -i\left[ \vecbb{H_0} + \vecbb{V} \right] L\right) = \vecbb{T}^{(0)}(L) + \vecbb{T}^{(1)}(L) + ...$. The lowest orders are $\vecbb{T}^{(0)}(L) = \exp (-i\vecbb{H_0} L)$ and 
\begin{align}
 \vecbb{T}^{(1)}(L) = -i\int_0^L \exp (-i\vecbb{H_0} R) \vecbb{V} \exp (-i\vecbb{H_0} (L-R)) \ \text{d} R.
\end{align}
This perturbation expansion can easily be implemented by discretizing the integral, in which case it maps to the dynamics of light interacting with discrete scatterers, such as atoms coupled to a waveguide \cite{chang2018, hood2016}.

\begin{figure}[t]
\begin{center}
 \includegraphics[height=.2\textwidth]{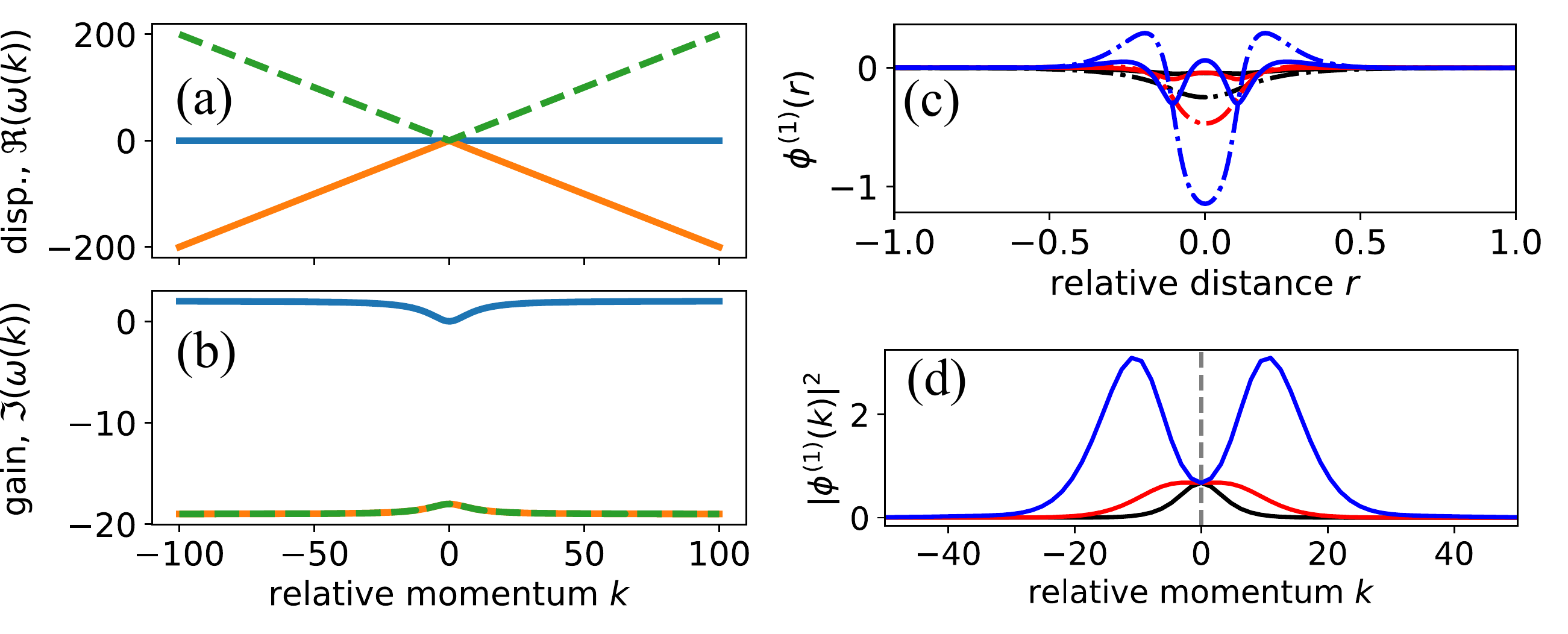}
\end{center}
\caption{For large loss ($\beta = 10$), the three polariton branches have quasi linear dispersion relations (a). Two of the branches are strongly damped, leaving a single active polariton (b). After a scattering event, the polariton wave's imaginary component (dashed dotted) develops a pronounced dip at $r=0$, while the real component (solid) plays a minor role, as illustrated for propagation distances $\Delta R=0$ (black), $\Delta R=1$ (red) and $\Delta R=2$ (blue) (c) at $R_b = 0.1$. In momentum space, the emergence of finite-time wings can be understood from the dominance of off-resonant wave components that are relatively protected from absorption (d). Color coding as in (c).}
\label{fig3}
\end{figure}

The first-order approximation to the scattering is excellent and reproduces the exact numerical results very well. Thus, we can understand the transmitted correlations as the superposition between the zeroth order wave and the many waves that are scattered at different points in the semiconductor. The initial state is an eigenstate of $\vecbb{H_0}$, such that it stays unaffected by $\vecbb{T}^{(0)}(L)$. The observed second-order photonic correlations functions (Fig.~\ref{fig2}(b)) thus find a simple interpretation: $\vecbb{V}$ projects the first-order scattered waves into the state $ep_+$, which then pick up a phase shift $\pi$ upon emission, and add up coherently during the propagation \cite{prasad2020}. They then interfere with the uncorrelated zeroth-order wave, resulting in destructive interference and the observed antibunched light feature in $g^{(2)}$. With increasing propagation distance, the amplitude of the first-order wave components at $\tau=0$ cancels the zeroth order weight and gives the special point $g^{(2)}(0)\approx 0$. We note that this cancellation is not perfect as the scattered wave amplitudes carry an imaginary component. Beyond the point of minimal $g^{(2)}(0)$, the same scattered wave functions dominate and ultimately lead to the bunching feature in $g^{(2)}(0)$.

If the loss rate is large ($\beta \gg 1$), qualitatively different photon correlations emerge in transmission, cf. Fig.~\ref{fig1} and Fig.~\ref{fig2}(c). In this case, the photonic correlation function also develops a dip after short propagation distances but its features are much narrower, given now by the blockade radius $R_b$, as the coherence length is now comparable or even smaller than the blockade radius. In contrast to the case of low $\beta$, the correlations do not spread significantly as the excitations propagate but retain a length scale of $R_b$. This can be explained by a single excitation blocking the creation of another exciton. A second photon will thus move faster until it eventually slows down again when it is a blockade radius away. Furthermore, the depression in $g^{(2)}$ still continues to the critical point of $g^{2}(0) \approx 0$ but additionally forms wings at finite time delay $\tau$, shortly before $g^{(2)}(0)$ begins growing again. The resulting photon correlations for reasonably long crystals still feature a significant reduction from the classical limit of $g^{(2)}(0)\geq1$, indicating the quantum nature of the transmitted light. For even larger $\beta$, the minimal value of $g^{(2)}(0)$ grows and the mentioned wings followed by finite-time bunching develop earlier. For very long crystals, the bunching feature at \textit{finite} $\tau$ dominates the correlations. 

A convenient picture for the lossy regime is given by the eigenstates of $\vecbb{H_0}$, the ``polariton basis''. Each of the three polariton branches has a dispersion and an associated damping rate (Fig.~\ref{fig3}(a-b)). While the dispersion relations are almost independent of the loss rate $\beta$, the damping rate of two polariton branches is proportional to $\beta$. For large $\beta$, these branches thus do not appreciably contribute to the dynamics and we can formulate an effective description in terms of the remaining polariton branch, $\vec{\phi}(R)$, alone: $i\partial_R \vec{\phi}(R) = \omega(k) \vec{\phi}(R) + \tilde{\vecbb{V}}\vecbb{\phi}(R)$, with the complex polariton branch $\omega(k)$ and the projected interaction matrix $\tilde{\vecbb{V}}$. Much like in the above case of low absorption, a scattering event induces a phase shift, leading to interference with the unperturbed wave component and, ultimately, the reduction of $g^{(2)}(0)<1$. Fig.~\ref{fig3}(c) shows this evolution of a single complex partial scattered polariton wave, $\phi^{(1)}$, as it propagates through the crystal after the scattering event. Conserving overall momentum, the interaction spreads the initial polariton wave from $k=0$ into a range of relative momenta (Fig.~\ref{fig3}(d)). The scattered polariton pairs thus contain a red- and a blue-shifted polariton each, which are both detuned from the exciton resonance and thus enjoy relative protection from absorption. These frequency components, whose characteristic length scale is given by the inverse blockade radius $R_b^{-1}$, hence experience relative gain, and eventually dominate the transmitted wave. The competition between interference and momentum-dependent damping results in the initial dip in $g^{(2)}(0)$ shown in Fig.~\ref{fig2}(c), followed by the dominance of photon pairs with finite separations $\tau \sim R_b$ after long propagation lengths.

\begin{figure}[t]
\begin{center}
  \includegraphics[height=.2\textwidth]{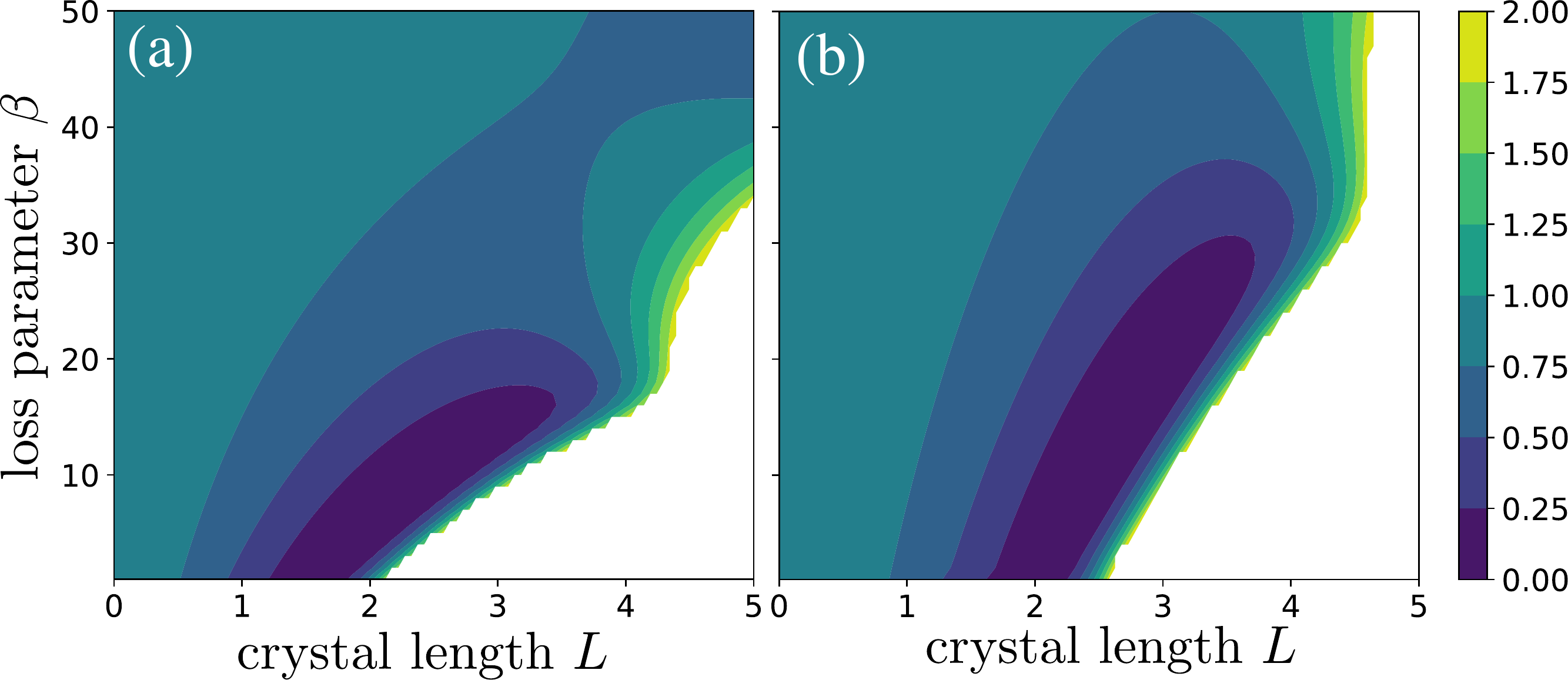}
\end{center}
\caption{The second-order correlation function at zero time delay, $g^{(2)}(0)<1$, indicates quantum light for a wide range of parameters at corresponding optimal crystal lengths $L$. The blockade radii are $R_b = 0.1$ and $R_b = 0.05$ in (a) and (b), respectively. All values $g^{(2)}(0)>2$ are whitened out for better visibility.}
\label{fig4}
\end{figure}

Photonic correlations are particularly interesting when they cannot originate from a classical state of light, as evidenced by the criterion $g^{2}(0)<1$ \cite{walls2007quantum, chavezmackay2020}. It is, therefore, an important question under which conditions ``quantum light'' can be produced by Rydberg excitons under the influence of loss. Fig.~\ref{fig4} shows a map of $g^{2}(0)$ as a function of the detuning $\beta$ and the crystal length $L$ for resonant excitation conditions. We observe that for each loss parameter $\beta$ there is an optimal range of crystal lengths at which the photonic state is most nonclassical. Lower values of $\beta$ are, in general, helpful for the emergence of quantum light. Remarkably, nonclassical light can, however, be found even in the limit of very large loss, $\beta \gg 1$. Smaller blockade radii can enhance the light's quantum features, although the required crystal lengths tend to be longer. This phenomenon can be traced back to the wider spread of momentum states that are populated by the scattering interaction, leading to rapidly dephasing wave function components and, hence, a quickly dropping $ee^{(1)}(r=0)$.

In conclusion, we have demonstrated that quantum light can be expected from the propagation of weak coherent light through a semiconductor crystal under a wide span of parameters. The excitonic field absorption lengths of cuprous oxide Rydberg states range from tens to hundreds of microns \cite{kazimierczuk2014giant}. The blockade radii have been measured to be several microns, justifying the assumed ratios $R_b/l_\text{abs} \lesssim 10\%$. Remarkably, the presence of a parallel absorption pathway, as given by a phonon-assisted background, is insignificant for the photon correlations in transmission, although it will increase the time to measure the predicted correlations \cite{kusmierek2022}. Nonradiative decay, on the other hand, has a strong bearing and tends to suppress quantum states of light. However, even for dissipation ratios as large as $\beta = 40$, quantum light can be obtained. These photonic correlations can be seen as unambiguous signatures of the Rydberg states and be used to benchmark their interactions. Furthermore, this work formalizes the link between atom-coupled waveguides and Rydberg photonics by presenting an effective perturbation description applicable to both systems. This offers an exciting alternative platform for quantum photonics as the easily controllable crystal length takes the place of the difficult-to-control number of atoms at the waveguide, the main cause of imperfect antibunching \cite{prasad2020}.  

The underlying mechanism in the present work is the modulation of absorption. Exciting outlooks therefore include using loss-mitigating schemes, that correlate photons through dynamical changes in the dispersion. The simplest such example is the despersive limit ($\Delta>\gamma$), as discussed in Appendix~\ref{sec:dispersive_regime}. There, the dynamics becomes fully coherent and can be described by an effective Schr\"odinger equation for the correlated motion of the photon pair through the crystal with an effective mass and a photon potential. Other promising avenues to explore are electromagnetically induced transparency \cite{waltherphonon2020} and cavity polaritons \cite{orfanakis2022, munjoz2019, delteil2019}. Under such conditions, it could be possible to observe more exotic effects, such as photon bound states \cite{maghrebi2015, cantu2020, PhysRevA.90.053804, liang2018}, Efimov states \cite{gullans2017} and many-body states \cite{bienas2020, iversen2021, iversen2022} including photon trains \cite{pohl2010, moos2015, otterbach2013} and topological states \cite{Clark2020, umucalilar2021}.

\begin{acknowledgments}
We thank Bj\"{o}rn Schrinski and Yuxiang Zhang for helpful discussions. V. W. thanks the Niels Bohr Institute for its hospitality for two visits, during which most of the work has been completed. This work has been supported by the NSF through a grant for the Institute for Theoretical Atomic, Molecular, and Optical Physics at Harvard University and the Smithsonian Astrophysical Observatory. A. S. acknowledges support of Danmarks Grundforskningsfond (DNRF 139, Hy-Q Center for Hybrid Quantum Systems).
\end{acknowledgments}

\clearpage
\onecolumngrid
\appendix

\section{Derivation details of the model} \label{sec:derivation_details}
The Hamiltonian $\mathcal{H}$ of the main text describes the coupled evolution of excitons and light as well as the exciton-exciton interactions. We first develop the description for a one-dimensional photon and generalize to three dimensions in the the next section. We assume the field outside the crystal to be a weak coherent cw-pulse with amplitude $\alpha$ which is approximated as a superposition of 0, 1 and 2 photons
\begin{align}
 |\Psi \rangle = e^{-\frac{|\alpha|^2}{2}} \sum_{n=0}^\infty \frac{\alpha^n}{\sqrt{n!}} |n\rangle \approx |0\rangle + \alpha |1\rangle + \frac{\alpha^2}{\sqrt{2}}|2\rangle,
\end{align}
where we have approximated $1-\frac{\alpha^2}{2} \approx 1$. The crucial approximation is to solve the quantum dynamics for 0, 1 and 2 excitations separately. In each sector, we can simply apply the effective Hamiltonian. In principle, this can be problematic since decay from the sector with $m$ excitations populates the sector with $m-1$. However, when considering the limit of very weak fields, $\alpha \ll 1$, where the amplitude of successively higher photon states is strongly suppressed, this effect can be neglected. The single-excitation subspace in the crystal reads
\begin{equation}
\begin{aligned}
 |\Psi(t) \rangle^{(1)} &= \int dz E(z,t) \hat{\mathcal{E}}^\dagger(z) |\emptyset\rangle 
 + \int dz P(z,t) \hat{\mathcal{P}}^\dagger(z) |\emptyset\rangle,
\end{aligned}
\end{equation}
while the two-photon sector has the general wave function 
\begin{equation}
\begin{aligned}
 |\Psi(t) \rangle^{(2)} &= \frac{1}{2} \int dz \int dz' \ EE(z,z',t) \hat{\mathcal{E}}^\dagger(z) \hat{\mathcal{E}}^\dagger(z') |\emptyset\rangle \\ 
 &+ \frac{1}{2} \int dz \int dz' \ PP(z,z',t) \hat{\mathcal{P}}^\dagger(z) \hat{\mathcal{P}}^\dagger(z') |\emptyset\rangle \\
 &+ \int dz \int dz' \ EP(z,z',t) \hat{\mathcal{E}}^\dagger(z) \hat{\mathcal{P}}^\dagger(z') |\emptyset\rangle. 
 \end{aligned}
\end{equation}
Because of the interchangeability of identical bosonic particles, we can in general assume $EE(z,z',t) = EE(z',z,t)$ and $PP(z,z',t)=PP(z',z,t)$.

We now formulate the Schr\"odinger equation by evaluating matrix elements and projecting onto the corresponding component, e.g.
\begin{align}
 \partial_t EE(z,z',t) =  \langle \emptyset | \hat{\mathcal{E}}(z) \hat{\mathcal{E}}(z') \hat{\mathcal{H}}|\Psi(t) \rangle.
\end{align}
We begin with the single-excitation component. As given in Eqs.~(2)-(3) of the main text, the equations of motion in the crystal are
\begin{align}
 \partial_t E(z,t) &= - v_g \partial_z E(z,t) - \gamma_\text{bg} E(z,t) - ig P(z, t) \\
 \partial_t P(z,t) &= - ig E(z, t) - \frac{\Gamma}{2} P(z, t),
\end{align}
where we introduce $\Gamma = \gamma - 2i \Delta$ with the total excitonic decay and dephasing rate $\gamma$ and the background absorption rate $\gamma_\text{bg}$. In this formulation, we consider a no-jump evolution of the wave function, which reproduces the exact evolution as the final state of the decay (vacuum) has no evolution \cite{Manzoni2018}. If both excitations are inside the semiconductor, the equations of motion read
\begin{align}
 \partial_t EE(z,z',t) &= -v_g (\partial_z + \partial_{z'}) EE(z,z',t) - 2\gamma_\text{bg}EE(z,z',t) - ig \left[ EP(z',z) + EP(z,z',t) \right] \\
 \partial_t EP(z,z',t) &= -v_g \partial_z EP(z,z',t) - \gamma_\text{bg} EP(z,z',t) -ig EE(z,z',t) - \frac{\Gamma}{2} EP(z,z',t) - ig PP(z,z',t) \\
 \partial_t PP(z,z',t) &= -ig \left[ EP(z,z',t) + EP(z',z,t) \right] - \Gamma PP(z,z',t) -i V(|z-z'|) PP(z,z',t),
\end{align}
with the interaction potential $V(R) = C_6/R^6$.

Here and in the following, we specialize to the steady state, where we can eliminate the component $PP(z,z')$. In the semiconductor, the single-excitation and the \textit{non-interacting} two-excitation sector are solved by (products of)
\begin{align}
 P(z) = - \frac{2ig}{\Gamma} E(z), \qquad E(z) = \alpha e^{- \left( \frac{2g^2}{v_g\Gamma} + \frac{\gamma_\text{bg}}{v_g} \right)z}.
\end{align}
Hence, as described in the main text, it is convenient to simplify the equations of motion by factoring out the non-interacting product functions $f_0(z,z') = g(z)g_1(z')$, viz by using the ansatz $F(z,z') = f(z,z')f_0(z,z')$. This yields the equations of motion 
\begin{equation}\label{eq:eom_time_independent}
\begin{aligned}
 (\partial_z + \partial_{z'}) ee(z,z') &= \frac{1}{1-2ix} \left[2 ee(z,z') - ep(z',z) - ep(z,z') \right] \\
 \partial_z ep(z,z') &= \beta (1-2ix) \left[ ee(z,z') - ep(z,z') \right] \\
 &  \ \ \ + \frac{1}{1-2ix} ep(z,z') - \frac{1}{2} \frac{1}{1-2ix + i \frac{\bar{C}_6}{|z-z'|^6}} \left[ ep(z,z') + ep(z',z) \right] ,
\end{aligned}
\end{equation}
where we defined the resonant excitonic field absorption length as the unit length, $l_\text{abs} = \frac{v_g\gamma}{2g^2}$ ($z \rightarrow z/l_\text{abs}$), the loss parameter $\beta = \frac{\gamma^2}{4g^2}$, the scaled detuning $x = \Delta/\gamma$ as well as $\bar{C}_6 = \frac{C_6}{\gamma l_\text{abs}^6} = \frac{64 g^{12} C_6}{v_g^6 \gamma^7}$. The blockade radius can then be defined as $R_b = \sqrt[6]{\bar{C}_6}$. This frame naturally provides particularly simple boundary conditions, given by $ee(z,z') = 1$ and $ep(z,z')=1$ when $z=0$ or $z'=0$. Finally, we move to center-of-mass coordinates, $R = \frac{z+z'}{2}$ and $r = z-z'$ and bring the equations into symmetrized form by defining $ep_{\pm}(R,r) = ep(R,r) \pm ep(R,-r)$ so that
\begin{align} \label{eq:coupled_equations_relative3}
&\partial_R \begin{pmatrix}
  ee(R,r) \\
  ep_+ (R,r) \\
  ep_- (R,r)
 \end{pmatrix} =
 \begin{pmatrix}
  0 & 0 & 0 \\
  0 & 0 & -2\partial_r \\
  0 & -2\partial_r & 0
 \end{pmatrix}\begin{pmatrix}
  ee(R,r) \\
  ep_+ (R,r) \\
  ep_- (R,r)
 \end{pmatrix} \\ +
 &\begin{pmatrix}
  2\frac{1}{1-2ix} & -\frac{1}{1-2ix} & 0 \\
  4(1-2ix)\beta & 2\left(-(1-2ix)\beta -iV_\text{eff}(r) \right) & 0 \\
  0 & 0 & 2\left(-(1-2ix) \beta + \frac{1}{1-2ix} \right)
 \end{pmatrix} 
 \begin{pmatrix}
  ee(R,r) \\
  ep_+ (R,r) \\
  ep_- (R,r)
 \end{pmatrix} = 0 \nonumber
\end{align}
with the effective potential 
\begin{align}
 V_\text{eff}(r) = -\frac{1}{2x+i} + \frac{1}{2x - \frac{R_b^6}{r^6}+i},
\end{align}
as given in Eqs.~(5)-(6) in the main text.

\section{Mode expansion of finite beams} \label{sec:mode_expansion}
So far, we have described the photon propagation in a one-dimensional channel. This description is accurate for single mode beams such as the $TEM_{00}$ mode considered in the main text, as long as the blockade radius exceeds the beam waist. Here, we illustrate effects when deviating from this limit.

We first recall that any solution to the free paraxial wave equation can be expanded in Hermite-Gauss modes. The general single-excitation expansion in modes labeled by quantum numbers $n$ and $m$ reads
\begin{align}
 \bar{f}(x,y,z) = \sum_{n,m} c_{n,m}(z) E_{n,m}(x,y,z)
\end{align}
with $E_{n,m}(x,y,z) = u_n(x,z) u_m(y,z) e^{ikz}$ and 
\begin{align}
 \int u^*_n(x,z) u_m(x,z) \ d x = \delta_{n,m},
\end{align}
independent of $z$. The primary use of this decomposition lies in the simple description of free propagation where $c_{n,m}(z) = c_{n,m}$, revealing the solution for all $z$ once it is known at one. With interactions, the equations for $n \neq m$ are coupled, as can be seen by the expansion in the complete basis
\begin{align} \label{eq:coupled_equations_relative_with_transverse}
&\partial_R \begin{pmatrix}
  (ee)^{mn}_{m'n'}(R,r) \\
  (ep_+)^{mn}_{m'n'} (R,r) \\
  (ep_-)^{mn}_{m'n'} (R,r)
 \end{pmatrix} =
 \begin{pmatrix}
  0 & 0 & 0 \\
  0 & 0 & -2\partial_r \\
  0 & -2\partial_r & 0
 \end{pmatrix}\begin{pmatrix}
  (ee)^{mn}_{m'n'}(R,r) \\
  (ep_+)^{mn}_{m'n'} (R,r) \\
  (ep_-)^{mn}_{m'n'} (R,r)
 \end{pmatrix} \\ +
 &\begin{pmatrix}
  2\frac{1}{1-2ix} & -\frac{1}{1-2ix} & 0 \\
  4(1-2ix)\beta & 2\left(-(1-2ix)\beta \right) & 0 \\
  0 & 0 & 2\left(-(1-2ix) \beta + \frac{1}{1-2ix} \right)
 \end{pmatrix} 
 \begin{pmatrix}
  (ee)^{mn}_{m'n'}(R,r) \\
  (ep_+)^{mn}_{m'n'} (R,r) \\
  (ep_-)^{mn}_{m'n'} (R,r)
 \end{pmatrix}  \nonumber \\ +
 &\sum_{\bar{m}\bar{n}\bar{m}'\bar{n}'}\begin{pmatrix}
  0 \\
  -i\tilde{V}^{mnm'n'}_{\bar{m}\bar{n}\bar{m}'\bar{n}'}(R,r) \cdot (ep_+)^{\bar{m}\bar{n}}_{\bar{m}'\bar{n}'} (R,r) \\
  0 
 \end{pmatrix}= 0. \nonumber
\end{align}
with the coupling matrix, here expressed as a function of $z,z'$ for convenience
\begin{equation}
\begin{aligned}
 \tilde{V}^{mnm'n'}_{\bar{m}\bar{n}\bar{m}'\bar{n}'}(z,z') = \int dx \int dy \int dx' \int dy' &u^*_{m}(x,z) u^*_{n}(y,z) u^*_{m'}(x',z') u^*_{n'}(y',z') \cdot \\ V_\text{eff}(x,y,z,x',z',z') \cdot 
 & u^*_{\bar{m}}(x,z) u^*_{\bar{n}}(y,z) u^*_{\bar{m}'}(x',z') u^*_{\bar{n}'}(y',z')
\end{aligned}
\end{equation}
Note that if the interactions are wider than the beam, $\tilde{V}$ becomes diagonal and we recover an decoupled system of equations as before. 

\begin{figure}[t]
 \begin{center}
 \includegraphics[height=.25\textwidth]{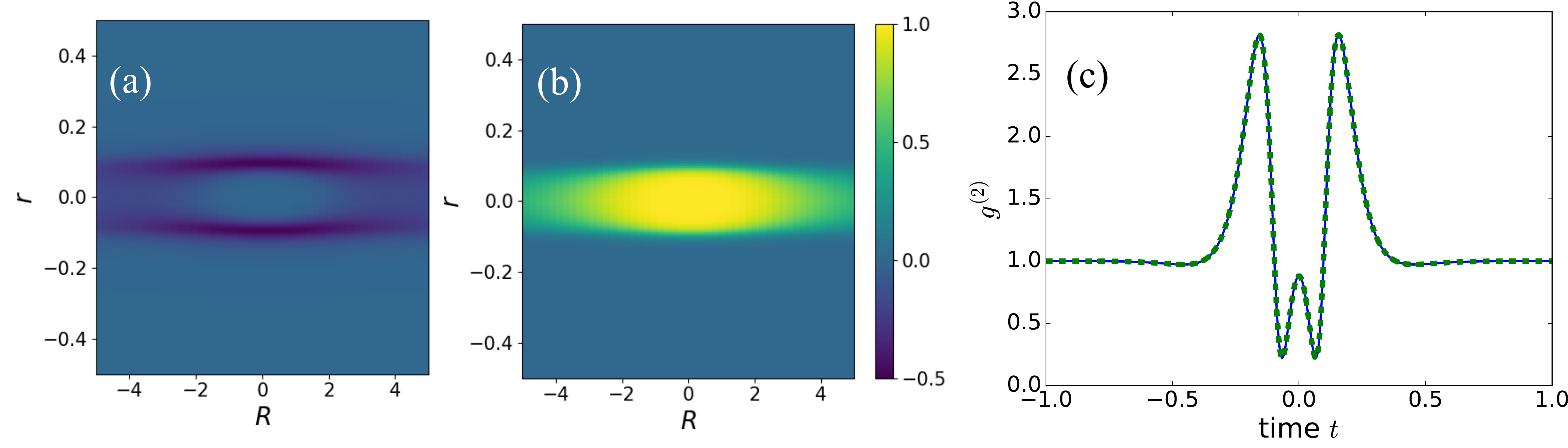}
 \end{center}
  \caption{The effective potential on resonance for a finite beam and finite-range interaction becomes a function of the center-of-mass position $R$ with a real (a) and an imaginary (b) component. (c) The resulting photonic correlation function at $R=3$ (solid line) is very close to the one-dimensional limit (dashed line) considered in the main text. The beam waist is set to be at the longitudinal center of the crystal. Parameters used are $\beta = 10$, $R_b = 0.1$ and beam waist $w_0 = 0.05$. The vacuum wavelength is $\lambda = 570$ nm, we approximate $v_g \approx \frac{c}{n}$ with the refractive index $n=2.7$ and we set the length scale to $l_\text{abs} = 92 \ \mu$m.}
\label{fig:finite_beam}
\end{figure}

While solving the full system is an interesting outlook, we consider only the input mode and assume it to be the $TEM_{00}$ mode which has the form
\begin{align}
 |u_0(x,z)|^2 = \sqrt{\frac{2}{\pi}} \frac{1}{w_0} \sqrt{\frac{z_R^2}{z^2+z_R^2}} \exp \left( -\frac{2}{w_0^2} \cdot \frac{z_R^2}{z^2+z_R^2} \cdot x^2 \right)
\end{align}
with the beam waist $w_0$ and the Rayleigh length $z_R$ determined by the dielectric constant $n$ and the (vacuum) wavelength $\lambda$
\begin{align}
 z_R = \frac{\pi w_0^2 n}{\lambda}. 
\end{align}
This gives a closed system of equations as before (with a modified potential) and physically corresponds to disregarding re-population of the initial mode. For weak scattering into the mode, this can be expected to be a good approximation. To simplify the computation (involving 4-dimensional integrals), we express the effective potential in cylindrical coordinates
\begin{align} 
 V_\text{eff} = -\frac{1}{2x+i} + \frac{1}{2x + i - \frac{R_b^6}{(r_\perp^2 + (r'_\perp)^2 - 2 r_\perp r'_\perp \cos(\phi_\perp - \phi'_\perp))^3}}.
\end{align}
Computationally, this allows to evaluate integrals separately by the following procedure
\begin{enumerate}
 \item fix $R_b$, $x$: carry out integrals over $\phi_\perp$ and $\phi'_\perp$
 \item fix $w_0$: carry out integrals over $r_\perp$ and $r'_\perp$ to obtain new effective potential $V_\text{eff}(r,R)$ with dependence on $R$
 \item fix $\beta$: solve photon propagation, using a combination of split-step (homogeneous part) and Euler integrator (potential).
\end{enumerate}
Fig.~\ref{fig:finite_beam}(c) shows a comparison between the beam propagation with in a Gaussian beam and the plain one-dimensional propagation ignoring any transverse propagation effects. The beam waist is chosen to lie in the center of the crystal. We see that the multimode effects are negligible at the displayed choice of $\beta=10$. The higher $\beta$ is, the longer crystals are required to observe significant effects. When the crystal becomes longer than $L\gtrsim 2 z_R$, the nonlinear effects are softened considerably. In the main text, we restrict our attention to parameters where this is not the case.

\section{Dispersive regime: effective Schr\"odinger equation} \label{sec:dispersive_regime}
\begin{figure}[h]
 \begin{center}
 \includegraphics[height=.4\textwidth]{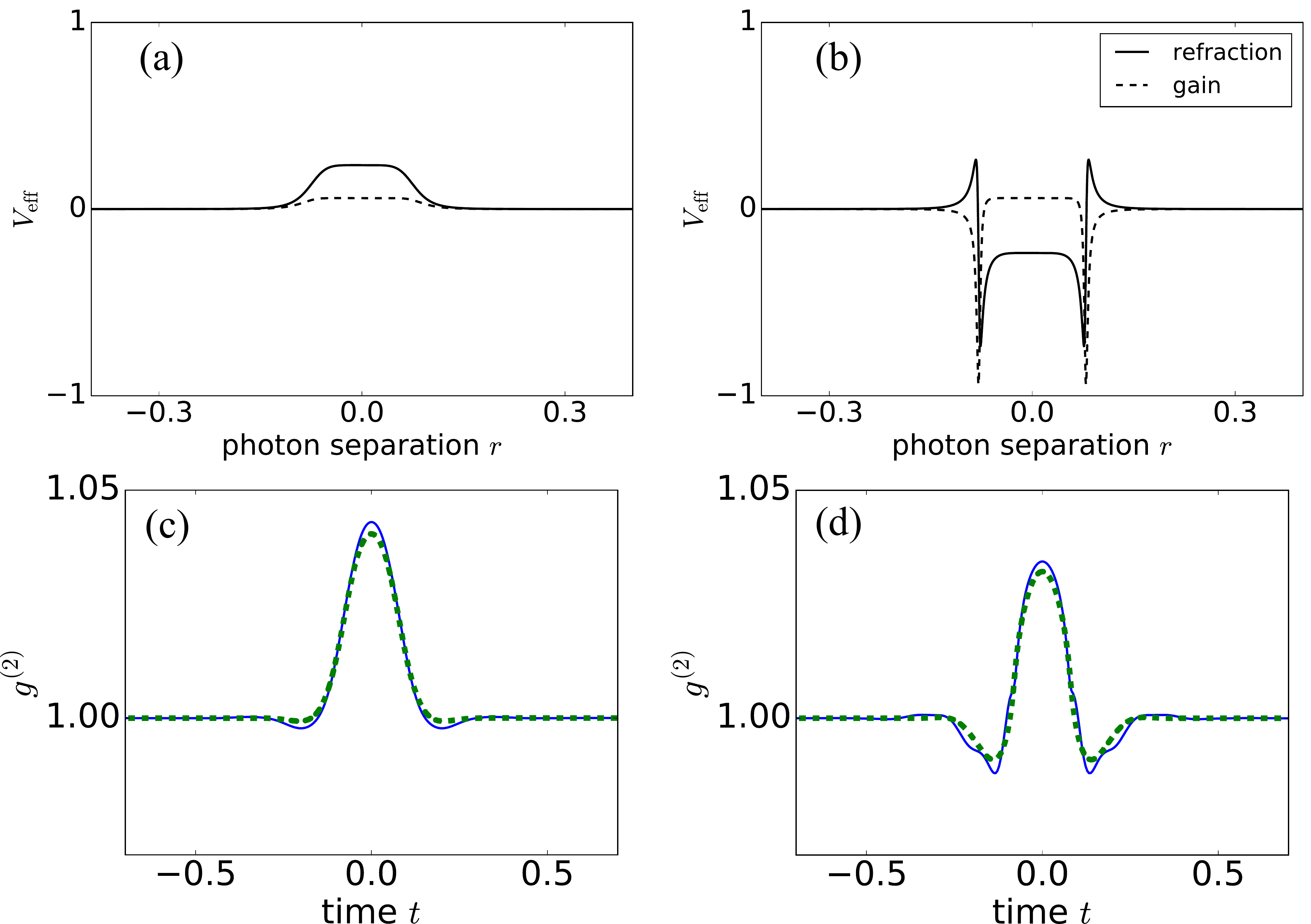}
 \end{center}
  \caption{The effective potential for $x=-2$ (a) is regular, while the potential at $x=2$ (b) shows a resonance (for repulsive interactions). Positive values of the imaginary part (dashed) of $V_\text{eff}$ indicate relative gain. Corresponding photon correlations after propagation through a crystal of length $L=10$ and loss parameter $\beta=10$ are shown in (c/d). Solid lines represent the exact solution, dashed lines depict the single-component approximation in Eq.~(\ref{eq:single_component}). }
\label{fig:fig2}
\end{figure}

The main text focuses on resonant excitation conditions, where the emergent photon correlations are strongest. Here, we outline the far-detuned limit, where the dynamics becomes more coherent and even permits a simplified description in terms of a Schr\"odinger equation. We begin by Fourier transforming Eq.~(\ref{eq:coupled_equations_relative3}) in the center-of-mass coordinate
\begin{align}
 ep_- (K,r) &= \frac{2}{2\left( -(1-2ix) \beta + \frac{1}{1-2ix} \right)  - iK}\partial_r ep_+ (K,r)  \\
 \frac{4}{2\left( -(1-2ix) \beta + \frac{1}{1-2ix} \right)  - iK}\partial^2_r ep_+(K,r) &= 4(1-2ix)\beta ee(K,r) + \left[ 2\left(-(1-2ix)\beta -i V_\text{eff}(r)\right) -iK \right] ep_+(K,r) \\
 ep_+(K,r) &= \left[ 2 - (1-2ix) iK\right] ee(K,r).
\end{align}
Eliminating the components $ep_-(K,r)$ and $ep_-(K,r)$ gives the \emph{exact} and \emph{general} single-component equation
\begin{equation}
\begin{aligned}
  \frac{4}{2\left( -(1-2ix) \beta + \frac{1}{1-2ix} \right)  - iK} \partial^2_r ee(K,r) -  \frac{4(1-2ix)\beta}{ 2 - (1-2ix) iK } ee(K,r) = \\
  \left[ 2\left(-(1-2ix)\beta -i V_\text{eff}(r)\right) -iK \right] ee(K,r).
\end{aligned}
\end{equation}
In the limit $2|x|\beta \gg K$, we can drop the dependence on the center of mass in the first term. To arrive at an intuitive equation, we further expand the second term in $K$
\begin{align}
 \frac{4(1-2ix)\beta}{ 2 - (1-2ix) iK } \approx 2(1-2ix)\beta - [(4x^2-1)i-4x]\beta K + \mathcal{O}(K^2),
\end{align}
an approximation we expect to work well for $xK \ll 1$. With these approximations, we can reorder the equation 
\begin{equation}
\begin{aligned}
   \left(iK - (1-2ix)^2 \beta i K \right)ee(K,r) \approx 
   -\frac{2}{ -(1-2ix) \beta + \frac{1}{1-2ix} } \partial^2_r ee(K,r)
 -2i V_\text{eff}(r) ee(K,r).
\end{aligned}
\end{equation}
and have thereby recovered a Schr\"odinger-like equation for the two-photon amplitude
\begin{equation}
\begin{aligned} \label{eq:single_component}
 \partial_R ee(R,r) \approx -\frac{2}{ \left[ -(1-2ix) \beta + \frac{1}{1-2ix} \right] \left(1-(1-2ix)^2 \beta \right) } \partial^2_r ee(R,r) - \frac{2iV_\text{eff}(r)}{1-(1-2ix)^2 \beta}  ee(R,r).
\end{aligned} 
\end{equation}
The resulting photon correlation are shown in Fig.~\ref{fig:fig2} and compared to the numerically exact result.
While the agreement is reasonably good already for $\beta =10$ and $|x|=2$, even more accurate results are obtained for larger values of both parameters. In the extremely dispersive limit, $1-2ix \approx 2ix$, the equation of motion becomes
\begin{equation}
\begin{aligned}
  i\partial_R ee(R,r) &\approx -\frac{1}{ 4x^3 \beta^2 } \partial^2_r ee(R,r) + \frac{V_\text{eff}(r)}{2x^2 \beta}  ee(R,r) \\
  &= -\frac{1}{ 4x^3 \beta^2 } \partial^2_r ee(R,r) + \frac{1}{4x^3 \beta} \left[ \frac{1}{1 -  \frac{1}{2x}\frac{R_b^6}{r^6}} -1 \right]  ee(R,r).
\end{aligned}
\end{equation}
Here, we can identify a ``mass'' $m = 2 x^3 \beta^2$ and a potential $U(r) = \frac{1}{4x^3 \beta} \left[ \frac{1}{1 -  \frac{1}{2x}\frac{R_b^6}{r^6}} -1 \right]$ that govern the pair-photon propagation.

\end{document}